\documentclass[5p]{elsarticle}
\usepackage{amssymb}
\usepackage{graphicx}
\journal{TIPP09 Proceedings in NIMA}

\begin{document}
\begin{frontmatter}
\title{Diamond Prototypes for the ATLAS SLHC Pixel Detector}
\author{Markus Cristinziani\fnref{fn1}}
\ead{cristinz@uni-bonn.de}
\address{Physikalisches Institut, Universit\"{a}t Bonn, Nussallee 12, 53115 Bonn, Germany}
\fntext[[fn1]{for the RD42 collaboration}
\begin{abstract}
Vertex detectors at future hadron colliders will need to cope with  
large particle fluences. Diamond is a particularly radiation hard material 
and exhibits further properties that makes it an attractive material for
such detectors. Within the RD42 collaboration several 
chemical vapor deposition diamond samples are being studied in the form of 
strip and pixel detectors. While the quality of the poly-crystalline diamond 
samples is constantly increasing and the feasibility of producing wafers has
been demonstrated, recently a single-crystal diamond pixel detector has been
assembled and characterized in a 100 GeV particle 
beam at CERN. Results on performance, detection efficiency, 
spatial resolution and charge collection are reported here together
with the latest radiation damage studies.
\end{abstract}
\end{frontmatter}
\section{Introduction}
While the Large Hadron Collider is scheduled to start collecting data this year, 
the design luminosity of the planned Super Large Hadron Collider (SLHC) of $10^{35}$cm$^{-2}$s$^{-1}$
poses a serious challenge for future vertex detectors close to the interaction region, 
in particular due to the harsh radiation environment.  In SLHC scenarios the total expected fluence 
at a radius of about 5cm will exceed $10^{16}$particles/cm$^2$. 
A number of studies are currently being performed to find solutions for detectors that can operate in such
environments. 
Chemical Vapor Deposition diamond, in either single-crystal (scCVD) or polycrystalline (pCVD) form,
is one material which is being considered for such detectors. Besides being radiation tolerant, diamond
is also an attractive material, since it is an excellent thermal conductor
and it exhibits no leakage current, fast collection times and a small input capacitance.

\begin{figure}[hbt] 
\begin{center} \includegraphics[height=0.44\columnwidth]{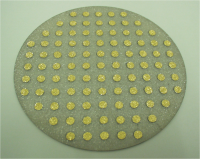}
\includegraphics[height=0.44\columnwidth]{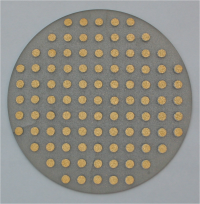}
\end{center} \caption{\label{fig:fig1} The growth side of two full 12cm diameter wafers metallised
with gold contacts 1cm apart.}
\end{figure}

Recently pCVD diamond material has become available in
the form of wafers. Figure~\ref{fig:fig1} shows two such diamond wafers
with as-grown collection distances of 315$\mu$m and 310$\mu$m respectively, produced by Element Six~\cite{elesix}.
The collection distance is a measure of the quality of
material and routinely exceeds 300$\mu$m.
After characterizing four full wafers, all indications are that
Element Six can reproducible grow and process high-quality detector-grade CVD diamond material.
Advances in the production of scCVD diamonds have lead to the production of sensors with sizes larger than 1cm$^2$ 
and thicknesses of up to 700$\mu$m. The production process is still in its development
and the sensors often come in irregular shapes. scCVD detector material is superior in terms of 
charge collection and homogeneity than poly-crystalline diamond sensors.

Several diamond pixel detectors have been constructed within the RD42 collaboration, developing and characterizing detectors,
in cooperation with the Fraunhofer Institute for Reliability and Microintegration (IZM)~\cite{izm}, 
performing fine-pitch flip-chip bump bonding. Single-chip devices and also 
multi-chip module detectors have been assembled and tested in particle beams. 
Testbeam results for a scCVD pixel detector~\cite{mathes} and a full module pCVD pixel detector~\cite{phd} are presented in this
paper, together with radiation tolerance studies~\cite{lhcc08}.

\section{Diamond pixel detectors}

\begin{figure}[hbt] 
\begin{center} \includegraphics[height=0.29\columnwidth]{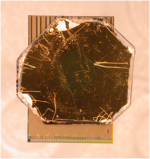}
\includegraphics[height=0.29\columnwidth]{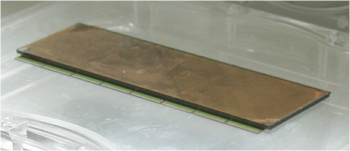}
\end{center} \caption{\label{fig:fig2} A scCVD single-chip (left) and a full-size pCVD diamond module connected with ATLAS pixel 
front-end chips through standard flip-chip bump-bonding.}
\end{figure}

Two diamond pixel detectors have been assembled with standard ATLAS pixel electronics, calibrated in the laboratory
and characterized in high-energy particle beams.
For assembly, both sides are metallised with Ti/W. One side is segmented into a $50\times400\mu$m$^2$ pixel
pattern to match the pattern of the ATLAS pixel front-end chip (FE-I3). 
Solder (PbSn) bumping and flip-chipping technology at IZM
has been employed to mate both parts. Figure~\ref{fig:fig2} shows the two assemblies, 
a single-chip detector, with an irregularly shaped $395\mu$m thick single-crystal diamond sensor (left) 
and a $800\mu$m thick full-size, $61 \times 16.5\mu$m$^2$ polycrystalline module with 16 front-end chips (right).

The ATLAS pixel electronics provides zero suppression in the readout with
a typical threshold level of 3000-4000 electrons and a noise of 160 electrons when used with silicon
detectors. As diamond sensors are basically free of leakage currents and have lower input
capacitance, the sensor was operated at thresholds of about 1700 electrons with a noise level of about
130 electrons, as determined from fitting the threshold curves of the pixels (S-curves). Analog
information is obtained via the in-pixel measurement of time-over-threshold (ToT) with an approximate
resolution of 8 bit. The ToT calibration is obtained for every pixel individually, by fitting the ToT response
to input pulses injected over a capacitance of 41.4fF.

\section{Testbeam performance}

The pCVD module has been tested in a 6 GeV electron beam at DESY. 
The measurements were affected by multiple scattering as
reported in~\cite{phd}. 
With a bias voltage of 800V a noise of
130 electrons at a threshold of $\sim 1500$ electrons were measured. More than $97\%$ of the 
electronics channels were operational. 

The scCVD single-chip module was tested in a 100 GeV pion test beam at CERN, using the 
Bonn-ATLAS telescope~\cite{BAT} as a reference for tracking and efficiency measurements.
It consists of double-sided silicon microstrip detectors with $50\mu$m pitch. Only tracks with a 
single hit in each telescope plane are selected. The extrapolated spatial uncertainty at the
device under test (DUT) is $5\mu$m. 


Unlike in silicon the electric field inside good scCVD diamond is fairly uniform and 
thus the spatial response of the detector was uniform for bias voltages larger than 
0.25 V/$\mu$m. Due to the faster charge collection smaller cluster sizes are observed
at increasing bias voltages. The measured most probable charge value (MPV) in the sensor at
400 V is 13.1 ke compatible with the expected value using the ToT charge
calibration outlined in~\cite{mathes}. The measured MPV increases with the applied bias
voltage and saturate above 100 V, indicating that the charge collection is complete
above these values. For full charge collection single and double hit clusters have 
roughly equal sharing. The overall efficiency to find a hit near a track point extrapolated
on the plane of the DUT is found to be $99.9\% \pm 0.1\%$.

\begin{figure}[hbt] 
\begin{center} \includegraphics[width=0.57\columnwidth]{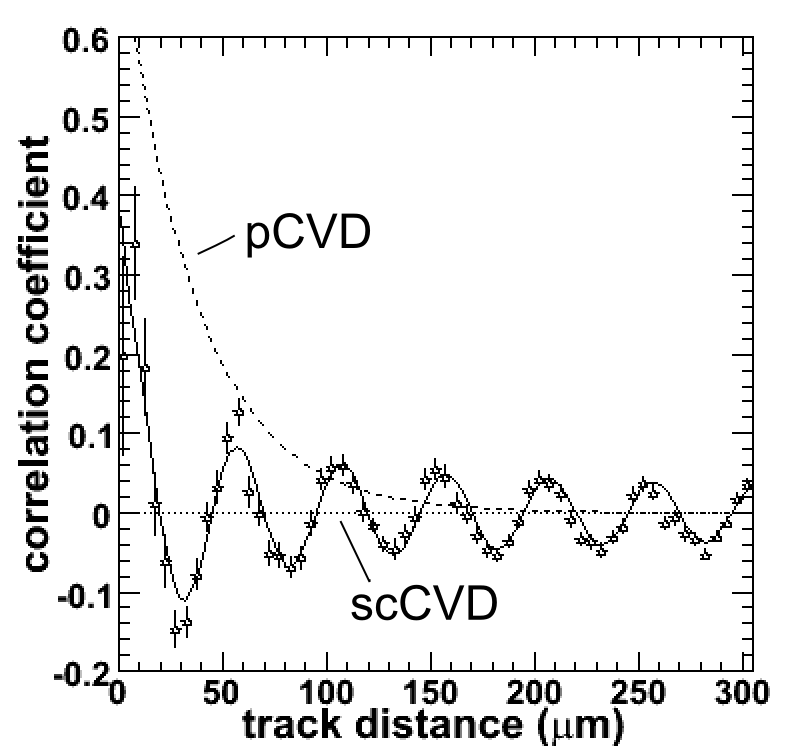}
\end{center} \caption{\label{fig:fig3} Correlation coefficient as a function of the track distance
between the tracks in two events. The lines are fits to the data in~\cite{lari} and~\cite{mathes} 
with a functional form of an exponential term plus a damped sinusoidal, describing the pixel pitch of $50 \mu$m.}
\end{figure}

Charge collection and space point reconstruction in pCVD are affected by the grain structure
and the trapping of electric charges at their boundaries which may cause horizontal polarization fields.
In~\cite{lari} it was observed that the distribution of residuals across the sensor is not
statistically distributed but exhibits regions with systematically positive or negative residuals
on a scale smaller than $100 \mu$m.
To quantify this apparent clustering, the linear correlation coefficient between the residuals
of all track pairs has been determined. Since each selected event has exactly one reconstructed track,
pairs are formed considering tracks of two different events. The average correlation coefficients are determined
in bins of the distance between the two tracks of a pair. An empirical function, a sum of 
a sinusoidal modulation with the periodicity of the pixel pitch with a falling exponential that accounts
for the residual shifts introduced by the grain structure, was found to provide a good description of the data.
In~\cite{lari} the data for a low-quality pCVD diamond were fit, resulting in a correlation length (the decay 
factor in the exponential) of $r_c= 36 \mu$m.
Here, we repeat this measurement for the \mbox{scCVD} single-chip detector. 
Since there are no grain boundaries the device exhibits optimal charge collection and a correlation length 
compatible with $0 \mu$m is observed~\cite{mathes}, indicating the absence of horizontal polarization fields.
This is illustrated in Figure~\ref{fig:fig3}.

\begin{figure}[hbt] 
\begin{center} \includegraphics[width=0.98\columnwidth]{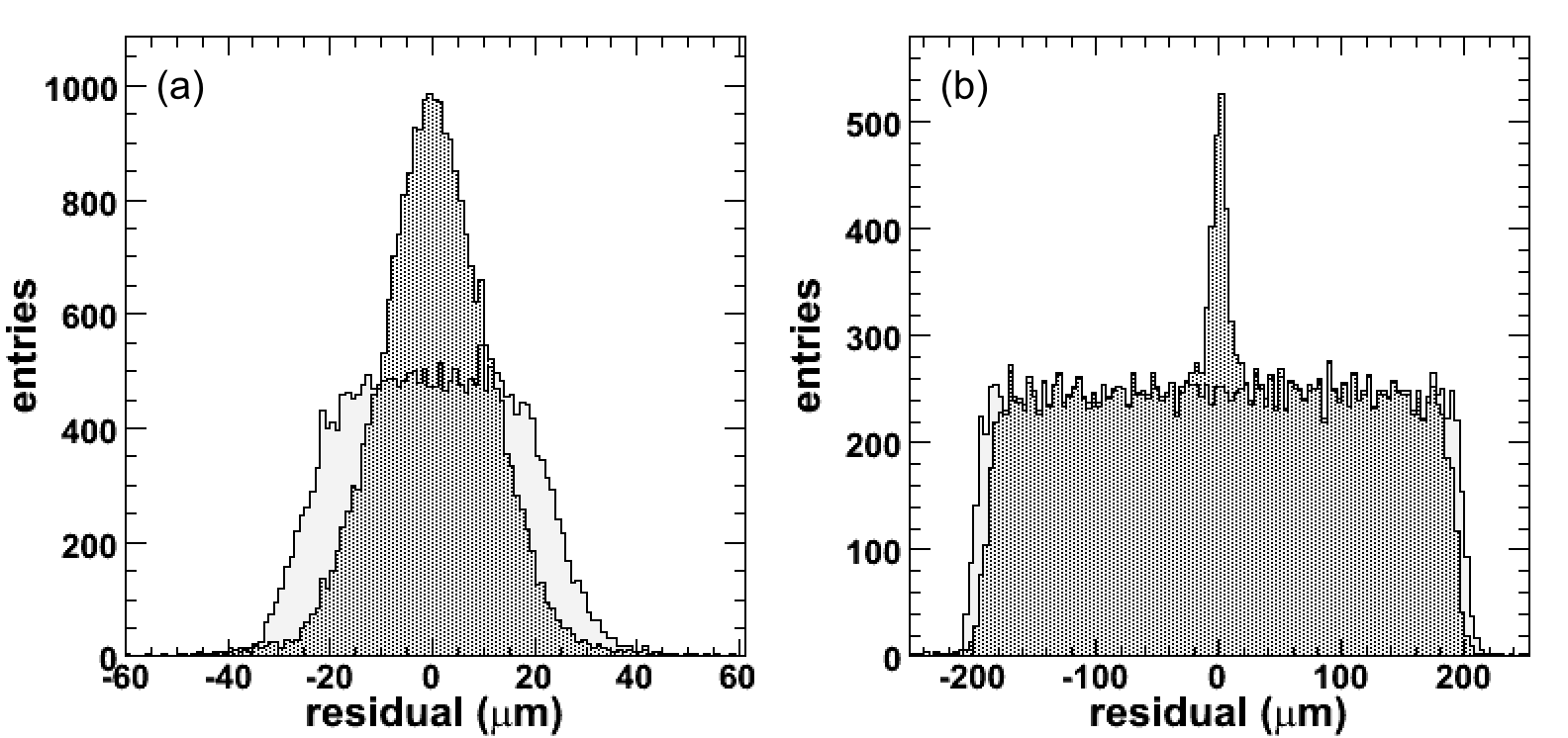}
\end{center} \caption{\label{fig:fig4} Spatial resolution of the scCVD device measured with respect
to the reference telescope, using digital information only (light shaded histograms) and analog
information via the $\eta$-algorithm (dark histograms). Bias voltage was 200V. Plotted is the difference between
the telescope space point and the measured space point. (a) resolution in the 50$\mu$m direction, (b) resolution in the 
400$\mu$m direction.}
\end{figure}

The spatial resolution is shown in Figure~\ref{fig:fig4}, where
the difference between the track position predicted by the telescope on 
the plane of the DUT (the scCVD sensor) and the reconstructed
hit of the DUT device is shown for both directions of the pixel.
For the hit reconstruction first only the digital information is used
(light gray distribution), i.e. the pixel with the largest charge signal 
above threshold collected in a cone around the extrapolated
track position is taken as the hit pixel and its center is assumed to
be the reconstructed position.
The expected digital resolutions of pixel pitch divided by $\sqrt{12}$ 
in both directions, i.e. 14.4 $\mu$m and 115.5 $\mu$m, respectively, 
folded by the resolution of the track extrapolation of about 5 $\mu$m, is
observed. 
Position reconstruction which exploits the analog
information available via the ToT
readout is done by using the $\eta$-algorithm~\cite{belau}.
The corresponding residual distribution is shown by the dark histograms
in Figure~\ref{fig:fig4}. It shows a pronounced peak with good resolution for
clusters with two pixel hits,
for which the $\eta$-algorithm improves the resolution substantially. In
the short pixel direction
 where the fraction of two hit clusters is about 40\% the
intrinsic resolution of the device,
after quadratically subtracting the telescope extrapolation error of
5$\mu$m, becomes ($8.9 \pm 0.1$)$\mu$m
for a bias voltage of 200V. The error quoted is obtained from the fit
to the residual distribution.
Repeating the analysis on several data samples indicate
an additional systematic uncertainty in the order of 0.5$\mu$m.
In the direction of the long pixel dimension, the resolution for events with
two-hit clusters ($\sim$10\%) is an order of magnitude better than for events with single hit clusters.
The space resolution depends on the bias voltage.
The charge collection efficiency and hence the total collected charge are field dependent
until full charge collection is obtained. Also the sharing of charge between pixels due to diffusion
depends on the field inside the sensor. The measured resolution shows an optimum between both
effects at bias voltages of 200V~\cite{mathes} for this device.

\section{Radiation Hardness}

The RD42 irradiation program consists of testing each sample prepared as a detector in a CERN test beam 
before and after irradiation. Previously, diamond samples were irradiated with reactor neutrons and charged
pions. Most recently we irradiated the highest quality
polycrystalline CVD diamond with 24 GeV protons from CERN PS to a fluence of $1.8 \cdot 10^{16}$ p/cm$^2$. 
\begin{figure}[hbt] 
\begin{center} \includegraphics[width=0.95\columnwidth]{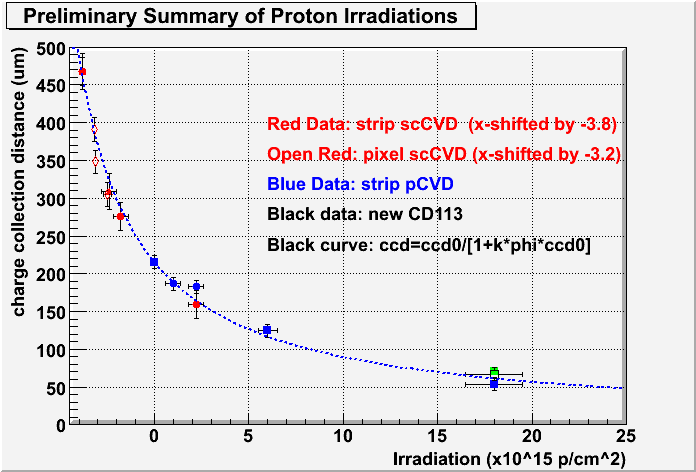}
\end{center} \caption{\label{fig:fig6} Summary of proton irradiation results for pCVD (blue points) and
scCVD (red points) material at an electric field of 1V/$\mu$m and 2V/$\mu$m (green square). The black curve 
is a standard damage curve 1/ccd = 1/ccd$_0$ + k$\phi$. With a shift of the scCVD doses of $-3.8 \cdot 10^{15}$ p/cm$^2$
the scCVD and pCVD data fall on a single curve indicating the damage due to irradiation is common to both.}
\end{figure}
Figure~\ref{fig:fig6} shows the signals observed in various pCVD and scCVD diamond sensors
after a range of irradiations. The scCVD diamond is expected to be representative of the next generation
high quality polycrystalline material. A single damage curve given by the equation ccd$^{-1}$ = ccd$_0^{-1} + k \phi$,
with ccd$_0$ the initial collection distance and $k$ the damage constant
can accommodate for the different measurements shown, with one universal constant $k$.
In particular, also the measurements on the scCVD fall on the same curve if shifted by $-3.8 \cdot 10^{15}$p/cm$^2$.
The interpretation is that, owing to the intrinsic crystallite structure, the unirradiated pCVD material 
behaves as effectively having the same number of trapping centers as the scCVD material after a dose of $-3.8 \cdot 10^{15}$p/cm$^2$.
After ten years of SLHC running typical fluences of $2 \cdot 10^{16}$p/cm$^2$ are
anticipated for vertex detectors at 4cm. From Figure~\ref{fig:fig6} we see
that diamond sensors still retain an average signal of $\sim$2500 electrons after such fluences~\cite{trischuk}.

\end{document}